\documentclass{aa}
\usepackage{times}
\usepackage{natbib}
\usepackage[dvips]{graphicx}
\bibpunct[]{(}{)}{;}{a}{}{,}

\def\xper{X Persei}
\def\XMM{{\em XMM--Newton}}
\def\EPIC{{\em EPIC}}
\def\MOS{{\em MOS}}
\def\Mone{{\em MOS1}}
\def\Mtwo{{\em MOS2}}
\def\pn{{\em pn}}
\def\XTE{{\em RossiXTE}}
\def\Ginga{{\em Ginga}}
\def\ASM{{\em ASM}}
\def\RX{RX J0146.9+6121}
\def\SAX{{\em BeppoSAX}}
\def\TreA{{3A 0535+262}}

\begin{document}

\title{\XMM\ observation of the persistent Be/neutron--star system \xper\ at a high--luminosity level}

\author{N. La Palombara \& S. Mereghetti}

\institute{{INAF -- Istituto di Astrofisica Spaziale e Fisica Cosmica Milano,
via Bassini 15, I--20133 Milano, Italy}}

\offprints{N. La Palombara, nicola@iasf-milano.inaf.it}

\authorrunning{N. La Palombara \& S. Mereghetti}

\titlerunning{\XMM~observation of \xper}

\abstract{We report on the \XMM~observation of the HMXRB \xper, the prototype of the persistent and low--luminosity Be/neutron star pulsars, which was performed on February 2003. The source was detected at a luminosity level of $\sim1.4\times10^{35}$ erg s$^{-1}$, which is the highest level of the latest three decades. The pulsation period has increased up to 839.3 s, thus confirming the overall spin--down of the NS detected in the previous observations. The folded light--curve has a complex structure, with features not observed at lower luminosities, and shows a significant energy dependence. The spectral analysis reveals the presence of a significant excess at low energies over the main power--law spectral component, which can be described by a black--body spectrum of high temperature (kT$_{\rm BB} \sim$ 1.5 keV) and small emitting region (R$_{\rm BB}\sim$ 340 m); its properties are consistent with a polar--cap origin. Phase--resolved spectroscopy shows that the emission spectrum varies along the pulse period, but it is not possible to prove whether the thermal component is pulsed or not.
}

\maketitle
\vspace{-0.3 cm}
\section{Introduction}
The X-ray source 4U 0352+309 is the brightest and perhaps most famous member of the Be/NS HMXRB class, whose optical counterpart was identified with the star X Persei \citep{BraesMiley1972,vandenBergh1972,BrucatoKristian1972,Weisskopf+1984}. In this system the companion pulsar accretes material from the massive, low--velocity equatorial wind of the Be primary star and emits X--rays as the gravitational energy of the infalling material is converted to X--ray luminosity. 

From the optical point--of--view, \xper\ (HD 24534, HR 1209) is a bright and highly variable star (on time scales from minutes to years) in optical and infrared (IR) brightness (V = 6.1--6.8, K = 5.2--6.7) and in emission--line strength \citep{Mook+1974,Roche+1997}. When the star is particularly bright, its spectrum displays strong emission in  H$_{\alpha}$ and other Balmer lines which marks this clearly as a Be--type system; when the star is faint, the emission lines disappear, and it appears to be a normal early--type star \citep[see:][; and references therein]{Fabregat+1992,Roche+1997}. In the past, sometimes it showed also extreme, extended low states \citep{Norton+1991}; during these periods the H$_{\alpha}$ line profile changed from emission to absorption, accompanied by a decrease in the IR flux by over a magnitude \citep{Reynolds+1992,Roche+1993}.

Such a variability is commonly supposed to be caused by the formation and dissipation of a disk around the star, as in other Be stars. However, between 1994 and 1995 \citet{LiuHang2001} observed that the H$_{\alpha}$ EW continued to increase even during the optical and near-infrared fading. They found that, although both components (line and continuum) arise from the same circumstellar envelope, their emission peaks at different evolution stages of the envelope itself, which correspond to different distances from the central star. In this scenario, the observed behaviour of the two components is attributed to the ejection of matter from the Be star as consecutive expanding rings.

The spectral class of the underlying OB star has been estimated to be O9.5 IIIe by \citet{Slettebak1982}. More recently \citet{Lyubimkov+1997} have used both spectroscopic and photometric observations of the low--luminosity diskless phase of \xper\ to infer the characteristics of the visible component; their results suggest a star of B0 Ve type, which is likely to have a mass of $\sim$ 13--20 M$_{\odot}$ and a radius of 5--10 R$_{\odot}$. With a projected equatorial velocity $v$sin$i$ of $\sim$ 1200 km s$^{-1}$ \citep{Reynolds+1992,Lyubimkov+1997} the Be star is probably seen under a low or intermediate inclination angle.

Considering UV, optical and IR data taken in diskless and near--diskless states \citet{Telting+1998} modelled the stellar photosphere with T$_{\rm eff}$ = 31000 K and log($g$) = 4. Then they derived \textit{E(B-V)} = 0.39 and, assuming R$_*$ = 9 R$_{\odot}$, estimated the distance to \xper\ as 950 $\pm$ 200 pc. With a Galactic latitude of -17$^{\circ}$, a distance estimate of $\sim$ 1 kpc implies that \xper\ would be $\sim$ 300 pc from the Galactic plane. This is rather far off the plane for an early--type star. However, it is within a few degrees of the plane of Gould's Belt and the distance estimates for \xper\ and the inferred extent of Gould's Belt \citep[$\sim$ 600 pc; ][]{Torra+2000} are marginally compatible.

The X--ray source has an unusually low luminosity, since up to now it has varied on timescales of years between $\sim$ 2.5 and $\sim$ 15$\times 10^{34}$ ergs s$^{-1}$; moreover, while most Be/NS binary systems are transient in nature, 4U 0352+309 is a source of persistent X-ray emission. This is perhaps due to the relatively wide and nearly circular orbit of the system, whose complete ephemeris were determined by \citet{Delgado-Marti+2001}. They used measurements made with the \textit{All Sky Monitor} (\textit{ASM}) on \XTE\ to perform a pulse timing analyses of the 837 s pulsations and found strong evidence for the presence of orbital Doppler delays. In this way they inferred that the orbit is characterized by a period P$_{\rm orb}$ = 250.3 days, a projected semimajor axis of the neutron star $a_{x}$sin$i$ = 454 lt--s, a mass function $f(M)=1.61 M_{\odot}$ and an eccentricity \textit{e} = 0.11. The measured orbital parameters, together with the known properties of the classical Be star \xper\, imply a semimajor axis a = 1.8--2.2 AU and an orbital inclination $i \sim$ 26--33$^{\circ}$.

\citet{Pfahl+2002} proposed this object as the prototype for a new class of HMXRBs characterised by long orbital periods (P$_{\rm orb} >$ 30 days) and low eccentricities. These systems are so much wide that no tidal circularisation can have occurred after the supernova explosion that created the neutron star, therefore all induced eccentricity must be primordial. While high velocity kicks are necessary to explain the eccentric orbits of most BeX transients \citep{van_den_Heuvel_van_Paradijs1997}, the low--\textit{e} orbit of \xper\ and of the other systems of its class must have formed in a supernova explosion without kick.

Its 837 s pulsation period was discovered with the \textit{Uhuru} satellite \citep{White+1976,White+1977,Margon+1977} and is still one of the longest periods of any known accreting pulsar \citep{Bildsten+1997,Haberl+1998}. The pulse profile of this source is almost sinusoidal. Some authors \citep{Robba+1996,Coburn+2001} have reported the presence of structures in the hardness ratio (4--11 keV)/(1--4 keV), i.e., a sharp hardening episode corresponding to the pulse minimum and two softening episodes just before and after this hard spike.

Like many accreting pulsars, the history of the pulse period of 4U 0352+309 shows no evidence for a regular spin--up. For the first years after the discovery of pulsations, the pulsar exhibited apparently erratic pulse frequency variations on a time-scale of a few days, which were superposed on a long term trend in which it was spinning up at the rate of $\dot {\rm P}/{\rm P} = -1.5 \times 10^{-4}$ yr$^{-1}$. This was followed by a large apparent torque reversal around 1978; since then, although the spin period measurements have been infrequent, the star has been spinning down at an average rate of $\dot {\rm P}/{\rm P} = 1.3 \times 10^{-4}$ yr$^{-1}$ \citep[; and references therein]{White+1976,Robba+1996,DiSalvo+1998,Delgado-Marti+2001}. In all, there have been only about two dozen determinations of the pulse period of 4U 0352+309, so that its pulse period behavior on short timescales (e.g., weeks, months, and even a year) is not well documented.

The transitions between the spin-up and spin-down stages occur without any significant variations of the system X--ray luminosity and, therefore, cannot be associated with the transitions of the neutron star between the accretor and propeller states. The long--term spin--down trend argues against the presence of a persistent accretion disk in this system; in contrast, the assumption of a spherical geometry in the accretion flow allows us to interpret the spin period of the neutron star in \xper\ in terms of the equilibrium period, P$^{sph}_{eq}$, provided the average (on a time-scale of $>$ 20 yr) value of the wind velocity is 350--400 km s$^{-1}$. The spin--up/spin--down behaviour of the source within this scenario can be associated with variations of the stellar wind velocity, which are due to the activity of the massive component and the orbital motion of the neutron star, whose trajectory is inclined to the plane of decretion disk of the Be companion \citep{Ikhsanov2007}.

Another unusual aspect of 4U 0352+309 is its X--ray spectrum. Most accreting X--ray pulsar spectra are well described by a standard model, in which a simple power law with a photon index of $\sim$ 1 is exponentially cut off above $\sim$ 20 keV \citep{WhiteSwankHolt1983}. The source 4U 0352+309, on the other hand, does not follow this standard spectral shape; it is possible that this is due to the low mass accretion rate. As a result, a variety of different models have been used to fit the spectrum. Using observations taken with \XTE\, \citet{Coburn+2001} found that all these models were improved by the addition of a cyclotron resonant scattering feature (\textit{CRSF}) at $\sim$ 29 keV. In particular, the model that best fitted the data is a combination of a blackbody plus a power law modified by a \textit{CRSF} at 28.6 keV, which implies a magnetic field strength of 2.5(1+z)10$^{12}$ G in the scattering region (where \textit{z} is the gravitational redshift). In the phase--averaged spectrum there is no evidence of a fluorescent Fe K-shell emission line in the range 6.4--6.7 keV, and the phase--resolved analysis shows that the blackbody and cyclotron line energies are consistent with being constant through the pulse.

Recently, \citet{Grundstrom+2007} have observed that between 2000 and 2006 the circumstellar disk grew to near record proportions, and concurrently the system X--ray flux dramatically increased, presumably due to the enhanced mass accretion from the disk. The H$_{\alpha}$ emission equivalent widths were the largest measured over the last few decades \citep{Roche+1993,Piccioni+2000,Clark+2001} and they indicated that the disk had grown significantly in size. The apparent increase in X--ray flux that accompanied the disk expansion confirms that the X-ray source is powered by gas from the Be star disk. However, \citet{Grundstrom+2007} found that the largest H$_{\alpha}$ half-maximum emission radius of the circumstellar disk was still much smaller than the separation between the Be star and neutron star. Therefore they suggested that tidal forces at periastron may excite a two--armed spiral in the disk, and that mass transfer will occur predominantly after periastron as the gas in the arm extension is accreted by the neutron star. The observed phase of X--ray maximum, which occurs about one quarter of a period after periastron, is consistent with this picture.

\vspace{-0.3 cm}
\section{Observations and data reduction}\label{sec:2}

\xper~was observed by \XMM~for $\sim$ 31 ks on February 2$^{nd}$, 2003. The three \EPIC~cameras, i.e. one \pn\ \citep{Struder+2001} and two \MOS\ \citep{Turner+2001}, were active and operated in {\em Full Frame} mode. This gave time resolutions of 73 ms and 2.6 s, respectively. For all of them, the Medium thickness filter was used. While in the case of the two \MOS\ cameras the effective source exposure time is $\sim$ 31 ks, it is much lower ($\sim$ 16 ks) for the \pn\ camera, since it was affected by the `counting mode' effect during the observation.

We used  version 7.0 of the \XMM~{\em Science Analysis System}
({\em SAS}) to process the event files.  After the standard
pipeline processing, we looked for possible periods of
high instrumental background, due to flares of soft protons with
energies less than a few hundred keV. We found that the first
$\sim$ 20 ks of the observation were affected by a high soft--proton
contamination. However, since \xper~was detected with a very high count--rate
(larger than 3 cts s$^{-1}$ in each camera), much higher than that of the background, the
soft--protons during the bad time intervals have a negligible
effect on the source spectral and timing analysis (their count rate in the source extraction area is less than 0.1 cts s$^{-1}$). Hence for our analysis we used the data of the whole observation.

\vspace{-0.3 cm}
\section{Timing analysis}\label{timing}

Due to the very high count rate, the data collected by the three \EPIC\ cameras were highly affected by photon pile--up. For this reason, we had to exclude the central area of the source PSF and to consider only data in an annulus area.

For the \Mone\ camera we selected an extraction corona with internal and external radius of 45 and 70$''$, respectively; for the \Mtwo\ camera the corresponding parameters are 40 and 70$''$. In the case of the \pn\ camera the source was imaged close to a CCD gap, therefore we were forced to select a very narrow corona, with internal and external radius of 30 and 35$''$, respectively. We checked with the \textit{SAS} task \textit{epatplot} that, with these selections, no event pile--up affected our data, therefore we used them to accumulate the source light--curves. The background curves were accumulated on large circular areas with no sources and radius of 60$''$, 40$''$  and 45$''$ for the \Mone\, \Mtwo\ and \pn~camera, respectively.

\begin{figure}[h]
\centering
\resizebox{\hsize}{!}{\includegraphics[angle=-90,clip=true]{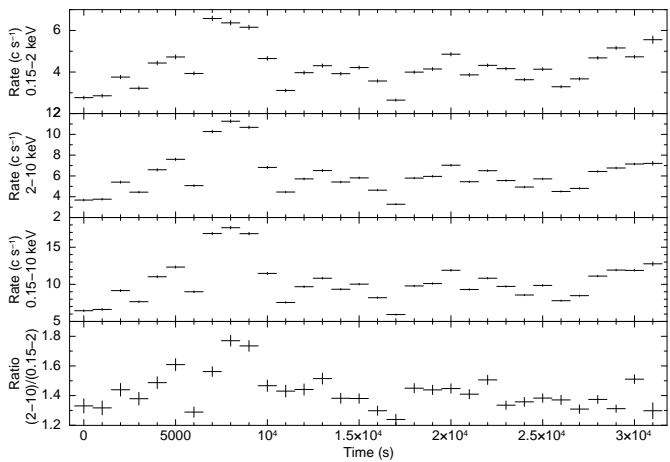}}
\caption{Background subtracted light curves of \xper~in the energy ranges 0.15--2, 2--10 and 0.15--10 keV, with a time bin of 1 ks.}
\label{lc_sum_1ks}
\end{figure}

\begin{figure}[h]
\centering
\resizebox{\hsize}{!}{\includegraphics[angle=-90,clip=true]{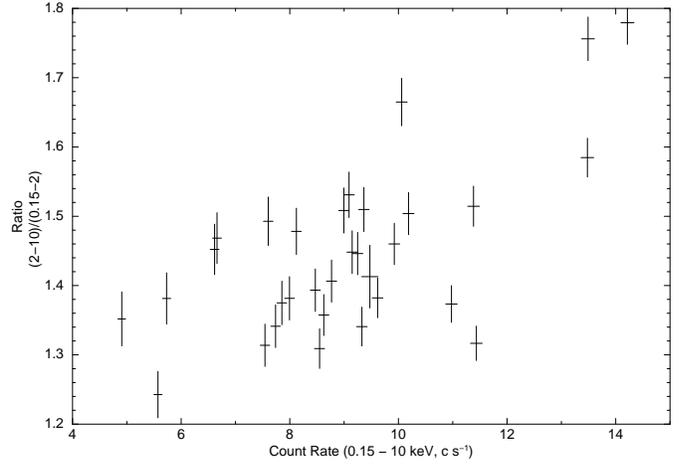}}
\caption{Hardness--ratio variation of \xper~as a function of the 0.15--10 keV count rate, with a time bin of 1 ks.}
\label{hr_cr_sum}
\vspace{-0.5 cm}
\end{figure}

In Fig.~\ref{lc_sum_1ks} we report the total background--subtracted light curves in the 0.15--10 keV energy range. It shows that, during the \XMM~observation, there were significant flux changes of the source on a few hours time-scale. Variations up to $\sim$ 50 \% around the average level of $\sim$ 10 cts s$^{-1}$ are evident. Moreover, the source \textit{hardness--ratio} (HR) between the energy bands above and below 2 keV varies up to $\sim$ 30 \% around the average value of 1.45 and, over long time-scales, increases with the total count rate (Fig.~\ref{hr_cr_sum}).

\begin{figure}[h]
\centering
\resizebox{\hsize}{!}{\includegraphics[angle=-90,clip=true]{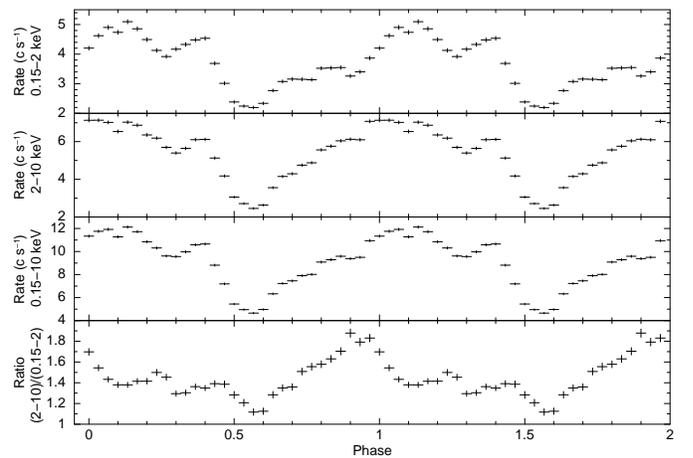}}
\caption{Background subtracted light curves of \xper~in the energy ranges 0.15--2, 2--10 and 0.15--10 keV, folded at the best--fit period.}
\label{flc_1}
\vspace{-0.5 cm}
\end{figure}

To obtain a  measure of the pulse period, we converted the
arrival times to the solar system barycenter and performed a folding
analysis using the source events of three cameras. In this way we derived a period
of 839.3 $\pm$ 0.3 s. In Fig.~\ref{flc_1} we show the folded light
curves in the three energy intervals 0.15--2, 2--10 and 0.15--10 keV, together with the folded HR between the hard and soft light curves. The pulse profile shows a main peak at phase $\sim$ 0--0.1, a secondary peak at phase $\sim$ 0.4 and a deep minimum at phase $\sim$ 0.5--0.6. The profile is energy dependent, because the main peak is at phase $\sim$ 0 and $\sim$ 0.1 in the high and low energy ranges, respectively. Moreover, below 2 keV it shows a plateau in the phase range 0.7--0.9, which has no counterpart above 2 keV. This energy dependence is confirmed by the HR profile, whose maximum is not aligned with the count--rate one. The same plot shows also that there is not a simple correlation between the hardness and the total count rate: we observe the same HR value at completely different count rate levels, but also very different HR values for the same count rate. This indicates that the spectral hardness of \xper~does not depend in a simple way on its flux level.

\begin{figure}[h]
\centering
\resizebox{\hsize}{!}{\includegraphics[angle=-90,clip=true]{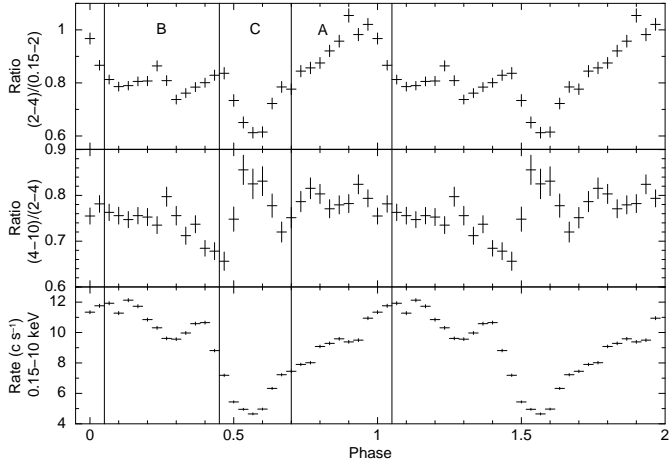}}
\caption{Hardness ratios of the three light curves of \xper\ in the energy ranges 0.15--2, 2--4 and 4--10 keV, as a function of the pulse phase. The vertical lines indicate the phase intervals used for the spectral analysis.}
\label{flc_2}
\vspace{-0.5 cm}
\end{figure}

In order to go deeper in our investigation, we have considered also a finer energy division by dividing the hard energy range in the two sub--ranges 2--4 and 4--10 keV. In Fig.~\ref{flc_2} we compare the folded \textit{soft} and \textit{hard} hardness ratios, i.e. the count--rates ratios in the energy bands (2--4)/(0.15--2) keV and (4--10)/(2--4) keV, respectively. In this way we can show that the minimum of the first ratio (and of the total light--curve) coincides with the maximum of the second one. This behaviour was already observed by \citet{Robba+1996} with the data collected in 1990 by the \textit{GINGA} satellite.

\vspace{-0.3 cm}
\section{Spectral analysis}\label{sec:3}
For the source and background spectra we considered the same extraction regions we used for the light curves. We accumulated all the events with pattern range 0--4 (i.e. mono-- and bi--pixel events) and 0--12 (i.e. from 1 to 4 pixel events) for the \pn~and the two \MOS~cameras, respectively.

We generated {\em ad hoc} response matrices and ancillary files using the {\em SAS} tasks {\em rmfgen} and {\em arfgen}. In order to ensure the applicability of the $\chi^{2}$ statistics, all spectra were rebinned with a minimum of 30 counts per bin; they were fitted in the energy range 0.3--10 keV using {\em XSPEC} 11.3.2. All spectral uncertainties and upper--limits are given at 90 \% confidence level for one interesting parameter.

\begin{figure}[h]
\centering
\resizebox{\hsize}{!}{\includegraphics[angle=-90,clip=true]{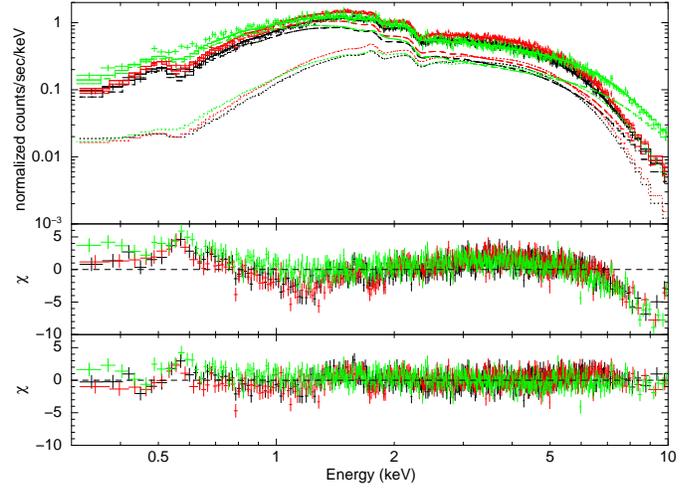}}
\caption{\textit{Top panel}: total spectrum of \xper~with the best--fit power--law (\textit{dashed line}) plus black--body (\textit{dotted line}) model. The spectrum of the \Mone, \Mtwo\ and \pn\ camera are shown in black, red and green, respectively. \textit{Middle panel}: residuals (in units of $\sigma$) between data and model in the case of the single power--law. \textit{Bottom panel}: residuals in the case of the power--law plus black--body.}\label{all_spetrum_powbb}
\vspace{-0.5 cm}
\end{figure}

After checking that separate fits of the three spectra gave consistent results, we fitted the spectra from the three cameras simultaneously in order to increase the count statistics and to reduce the uncertainties. The fit with an absorbed power--law yielded N$_{\rm H} =
(3.56\pm0.05)\times10^{21}$ cm$^{-2}$ and photon index
$\Gamma=1.17\pm0.01$, but with large residuals and
$\chi^{2}_{\nu}$/d.o.f. = 2.172/2131. The addition
of a black--body component improved  the fit quality significantly
(Fig.~\ref{all_spetrum_powbb}): we obtained N$_{\rm H} =
(2.86\pm0.10)\times10^{21}$ cm$^{-2}$,
$\Gamma=1.35\pm0.05$ and kT$_{\rm BB} =
1.35\pm0.03$ keV, with $\chi^{2}_{\nu}$/d.o.f. =
1.052/2129. The emission surface of the
thermal component has a radius R$_{\rm BB} = 392^{+11}_{-10}$ m
(for d = 1 kpc). The unabsorbed flux in the energy range 0.3--10
keV is $f_{\rm X}\sim1.2\times10^{-9}$ erg cm$^{-2}$ s$^{-1}$,
about 39 \% of which is due to the black--body component. Assuning a source distance of 1 kpc this flux implies an unabsorbed source luminosity $L_{\rm X}\sim1.4\times10^{35}$ erg s$^{-1}$.

We also attempted to fit the spectra by replacing the black--body component with alternative emission models, such as \textit{mekal}, thermal bremsstrahlung and broken power--law. In all these cases the results were worse than those obtained with the black--body model, since the reduced chi--squared was higher, the residuals were larger and/or the best--fit parameter values were unrealistic.

The spectra reported in Fig.~\ref{all_spetrum_powbb} show that the best--fit model still leaves an emission feature at 0.5--0.6 keV. We could fit it with a gaussian component at E = 0.58 keV, which has an equivalent width of 97 eV. On the other hand, we found no narrow Iron K$_{\alpha}$ lines between 6 and 7 keV, with an upper limit on its equivalent width of 0.1 keV at most. The final best--fit parameters are reported in Tab.~\ref{fit}.

\vspace{-0.5 cm}
\begin{table}[htbp]
\caption{Best--fit parameters for the total spectrum of \xper. The model parameters are the galactic interstellar absorption $A_{G}$, the main power--law component $PL$, the additional black--body component $BB$ and the Gaussian line $GL$.}\label{fit}
\begin{tabular}{ccc} \hline
Component 	& Parameter		& Value				\\ \hline
$A_{G}$		& N$_{\rm H}^{a}$	& 3.42$_{-0.08}^{+0.30}$	\\
$PL$		& $\Gamma$		& 1.48$_{-0.02}^{+0.02}$	\\
		& Flux @ 1 keV$^{b}$	& 9.48$_{-0.20}^{+0.64}$	\\
$BB$		& $kT$(keV)		& 1.42$_{-0.02}^{+0.04}$	\\
		& $R$(m)$^{c}$		& 361$_{-3}^{+3}$		\\
$GL$		& $E_{line}$(keV)	& 0.58$_{-0.04}^{+0.01}$	\\
		& $I_{line}^{d}$	& 2.13$_{-0.32}^{+0.22}$	\\
		& EQW (eV)		& 97$_{-15}^{+10}$		\\ \hline
d.o.f.		&			& 2126				\\
$\chi^{2}_{\nu}$&			& 1.021				\\ \hline
\end{tabular}
\begin{small}
\\
$^{a}$ $10^{21}$ cm$^{-2}$

$^{b}$ $10^{-2}$ ph cm$^{-2}$ s$^{-1}$ keV$^{-1}$

$^{c}$ for a source distance of 1 kpc

$^{d}$ $10^{-3}$ ph cm$^{-2}$ s$^{-1}$
\end{small}
\vspace{-0.5 cm}
\end{table}

\vspace{-0.3 cm}
\section{Phase--resolved spectroscopy}\label{spectroscopy}

Due to the hardness--ratio variations observed along the pulse--period, we performed a phase--resolved spectroscopy in order to study in more detail the source behavior. To this aim, we analysed the background subtracted spectra in the three phase intervals defined in Fig.~\ref{flc_2}, i.e. 0.7--1.05 (phase A), 0.05--0.45 (phase B), and 0.45--0.7 (phase C).

\begin{figure}[h]
\centering
\resizebox{\hsize}{!}{\includegraphics[angle=-90,clip=true]{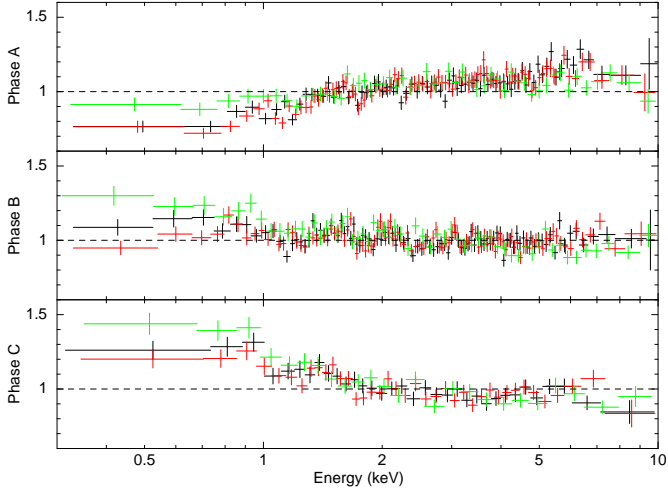}}
\caption{Ratios of the spectra corresponding to the phase intervals shown in Fig.~\ref{flc_2} to the average model spectrum. The parameters are fixed at the best--fit values for the total spectrum, except for the normalizations which are at their best--fit values for each phase interval.}\label{ratios}
\vspace{-0.5 cm}
\end{figure}

The first step was to fit all of them with the best--fit power law plus blackbody model of the phase averaged spectrum, leaving only the relative normalization factors free to vary. In Fig.~\ref{ratios} we report the ratios of the three spectra of each instrument to these renormalized average models. They show significant residuals for the spectra of phases A and C, which are harder and softer than the medium one, respectively. They clearly demonstrate the spectral variability as a function of the pulse phase.

We then fitted the three spectra independently. In all cases, the absorbed power--law model was not satisfactory, while the addition of a black--body component significantly improved the fit. Therefore we used this model for all the spectra, leaving all the parameters free to vary: the results are reported in Table~\ref{3spectra_fit}. They show that the absorbing column density is almost costant along the phase--pulse, as expected if it is due to the interstellar extinction. Moreover, both the power--law and the black--body components do not change significantly between phases A and B, neither in shape nor in intensity. Instead there is a relevant variation in phase C, as the power--law component steepens, the black--body temperature increases and the total luminosity reduces.

For completeness, we also looked for the presence, in the individual spectra, of the emission gaussian component at 0.58 keV which was found in the phase--averaged spectrum. We found that this line is well defined in the spectra of phases B and C, with an equivalent width of $\sim$ 105 and $\sim$ 75 eV, respectively; on the other hand, it is not well constrained in phase A.

\begin{table}[htbp]
\caption{Best--fit spectral parameters for the phase--resolved spectroscopy of \xper\, in the case of the independent fit of the three spectra. Errors are at a 90 \% confidence level for a single interesting parameter. N$_{\rm H}$, kT$_{\rm BB}$, and R$_{\rm BB}$ are measured in units of $10^{21}$ cm$^{-2}$, keV, and meters (for a source distance of 1 kpc), respectively.}\label{3spectra_fit}
\begin{tabular}{c|ccc} \hline
Spectral		&  \multicolumn{3}{c}{Phase Interval}								\\
Parameter		& A				& B				& C				\\ \hline
N$_{\rm H}$		& $3.23^{+0.19}_{-0.20}$	& $2.67^{+0.14}_{-0.14}$	& $2.94^{+0.28}_{-0.24}$	\\
$\Gamma$		& $1.30^{+0.09}_{-0.08}$	& $1.32^{+0.06}_{-0.06}$	& $1.61^{+0.17}_{-0.14}$	\\
kT$_{\rm BB}$		& $1.36^{+0.05}_{-0.05}$	& $1.28^{+0.05}_{-0.05}$	& $1.48^{+0.06}_{-0.08}$	\\
R$_{\rm BB}$		& 425				& 433				& 264				\\ \hline
$f_{\rm TOT}^{a}$	& 1.46				& 1.49				& 0.82				\\
$f_{\rm PL}^{a}$	& 0.83				& 0.97				& 0.49				\\
			& (57 \%)			& (65 \%)			& (60 \%)			\\
$f_{\rm BB}^{a}$	& 0.63				& 0.52				& 0.33				\\
			& (43 \%)			& (35 \%)			& (40 \%)			\\ \hline
$\chi^{2}_{\nu}$/d.o.f.	& 1.028/1465			& 1.032/1550			& 0.990/964			\\ \hline
\end{tabular}
\begin{small}
\\
$^{a}$ Unabsorbed flux in the energy range 0.3--10 keV, in units of $10^{-9}$ erg cm$^{-2}$ s$^{-1}$
\end{small}
\end{table}

In order to investigate the relative variations of the two
components with the period phase, we also simultaneously fitted
the three spectra forcing common values for N$_{\rm H}$, $\Gamma$ and kT$_{\rm BB}$.
In this case we obtained N$_{\rm H} = 2.94^{+0.09}_{-0.08}\times10^{21}$ cm$^{-2}$,
$\Gamma_{\rm PL}=1.40^{+0.04}_{-0.04}$ and kT$_{\rm BB} = 1.37^{+0.02}_{-0.02}$
keV, with $\chi^{2}_{\nu}$/d.o.f. = 1.105/3989; the corresponding
normalization values are reported in Table~\ref{3spectra_common}.
In this interpretation the spectral changes as a function of the
phase are reproduced by the variations in the relative
contribution of the two components. From this point of view, the power--law component provides the same flux during phases A and C, while it has a maximum during phase B. On the other hand, the contribution of the black--body component progressively decreases from phase A to phase C.

\begin{table}[htbp]
\caption{Best--fit values for the black--body and power--law normalizations, when the three spectra are fitted simultaneously with common values of N$_{\rm H}$ ($2.94^{+0.09}_{-0.08}\times10^{21}$ cm$^{-2}$), $\Gamma$ ($1.40^{+0.04}_{-0.04}$) and kT$_{\rm BB}$ ($1.37^{+0.02}_{-0.02}$ keV). Errors are at a 90 \% confidence level for a single interesting parameter}\label{3spectra_common}
\begin{tabular}{c|ccc} \hline
Spectral		& \multicolumn{3}{c}{Phase Interval}					\\
Parameter		& A			& B			& C			\\ \hline
I$_{\rm PL}^{a}$	& $70\pm3$		& $103\pm3$		& $67\pm2$		\\
R$_{\rm BB}^{b}$	& $465^{+11}_{-12}$	& $388^{+10}_{-11}$	& $247^{+10}_{-11}$	\\
$f_{\rm TOT}^{c}$	& 1.43			& 1.50			& 0.84			\\
$f_{\rm PL}^{c}$	& 0.65			& 0.96			& 0.62			\\
			& (46 \%)		& (64 \%)		& (74 \%)		\\
$f_{\rm BB}^{c}$	& 0.78			& 0.54			& 0.22			\\
			& (54 \%)		& (36 \%)		& (26 \%)		\\ \hline
\end{tabular}
\begin{small}
\\
$^{a}$ Intensity of the power--law component in units of $10^{-3}$ ph cm$^{-2}$ s$^{-1}$ keV$^{-1}$ at 1 keV

$^{b}$ Radius of the blackbody component (in metres) for a source distance of 1 kpc.

$^{c}$ Unabsorbed flux in the energy range 0.3--10 keV, in units of $10^{-9}$ erg cm$^{-2}$ s$^{-1}$
\end{small}
\vspace{-0.5 cm}
\end{table}

In order to confirm that the thermal component varies as a function of the rotational phase, we should prove that a constant black--body component is rejected by the data. To this aim, we modified the test model by imposing a common value of both kT$_{\rm BB}$ and R$_{\rm BB}$ for the three spectra, while both $\Gamma$ and I$_{\rm PL}$ could vary within them. The resulting best--fit has N$_{\rm H}= (3.19^{+0.08}_{-0.08})\times10^{21}$ cm$^{-2}$, kT$_{\rm BB} = 1.38^{+0.02}_{-0.02}$ keV and R$_{\rm BB}$ = $357^{+9}_{-9}$ m, while the power--law parameters are shown in Tab.~\ref{3spectra_BBcommon}. Also in this case the fit quality is very good ($\chi^{2}_{\nu}$/d.o.f. = 1.135/3989), comparable to the one obtained in the previous case. Moreover, in this case we find that during phase C most (i.e. $\sim$ 60 \%) of the total unabsorbed flux is due to the thermal component.

\begin{table}[htbp]
\caption{Best--fit values for the black--body and power--law normalizations, when the three spectra are fitted simultaneously with common values of N$_{\rm H}$ ($(3.19^{+0.08}_{-0.08})\times10^{21}$ cm$^{-2}$), kT$_{\rm BB}$ ($1.38^{+0.02}_{-0.02}$ keV), and R$_{\rm BB}$ ($376^{+9}_{-9}$ m for a source distance of 1 kpc). Errors are at a 90 \% confidence level for a single interesting parameter}\label{3spectra_BBcommon}
\begin{tabular}{c|ccc} \hline
Spectral		& \multicolumn{3}{c}{Phase Interval}								\\
Parameter		& A				& B				& C				\\ \hline
$\Gamma$		& $1.21^{+0.02}_{-0.03}$	& $1.44^{+0.03}_{-0.03}$	& $2.16^{+0.07}_{-0.07}$	\\
I$_{\rm PL}^{a}$	& $85^{+2}_{-3}$		& $112^{+3}_{-3}$		& $64^{+3}_{-2}$		\\
$f_{\rm TOT}^{b}$	& 1.51				& 1.51				& 0.85				\\
$f_{\rm PL}^{b}$	& 1.00				& 1.00				& 0.34				\\
			& (66 \%)			& (66 \%)			& (40 \%)			\\
$f_{\rm BB}^{b}$	& 0.51				& 0.51				& 0.51				\\
			& (34 \%)			& (34 \%)			& (60 \%)			\\ \hline
\end{tabular}
\begin{small}
\\
$^{a}$ Intensity of the power--law component in units of $10^{-3}$ ph cm$^{-2}$ s$^{-1}$ keV$^{-1}$ at 1 keV

$^{b}$ Unabsorbed flux in the energy range 0.3--10 keV, in units of $10^{-9}$ erg cm$^{-2}$ s$^{-1}$
\end{small}
\end{table}

\begin{figure}[h]
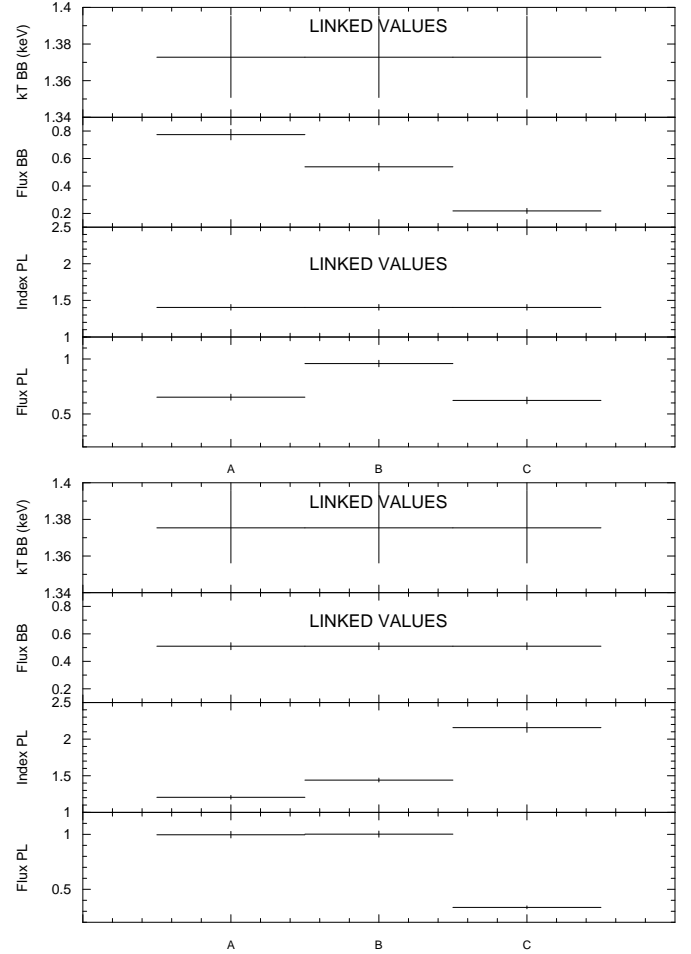

\centering
\resizebox{\hsize}{!}{\includegraphics[angle=-90]{normalizzazioni_comuneABCD_new.ps}}
\resizebox{\hsize}{!}{\includegraphics[angle=-90]{normalizzazioni_BBcomuneABCD_new.ps}}
\caption{Pulse--phase dependence of the black--body temperature, the power--law photon index and the unabsorbed flux of the two components (in units of 10$^{-9}$ erg cm$^{-2}$ s$^{-1}$), in the case of a common temperature and index (\textit{top}) and of a common black--body (\textit{bottom})}\label{both_3spectra}
\vspace{-0.5 cm}
\end{figure}

The above results are summarized in Fig.~\ref{both_3spectra},
where we show the pulse--phase dependence of the black--body
temperature, the power--law photon index and the unabsorbed flux
of the two components for both fits. Since they have a similar statistical
quality, we can neither confirm nor deny that the thermal
component is variable.

\vspace{-0.3 cm}
\section{Discussion}\label{sec:6}

\begin{figure}[h]
\centering
\resizebox{\hsize}{!}{\includegraphics[angle=0]{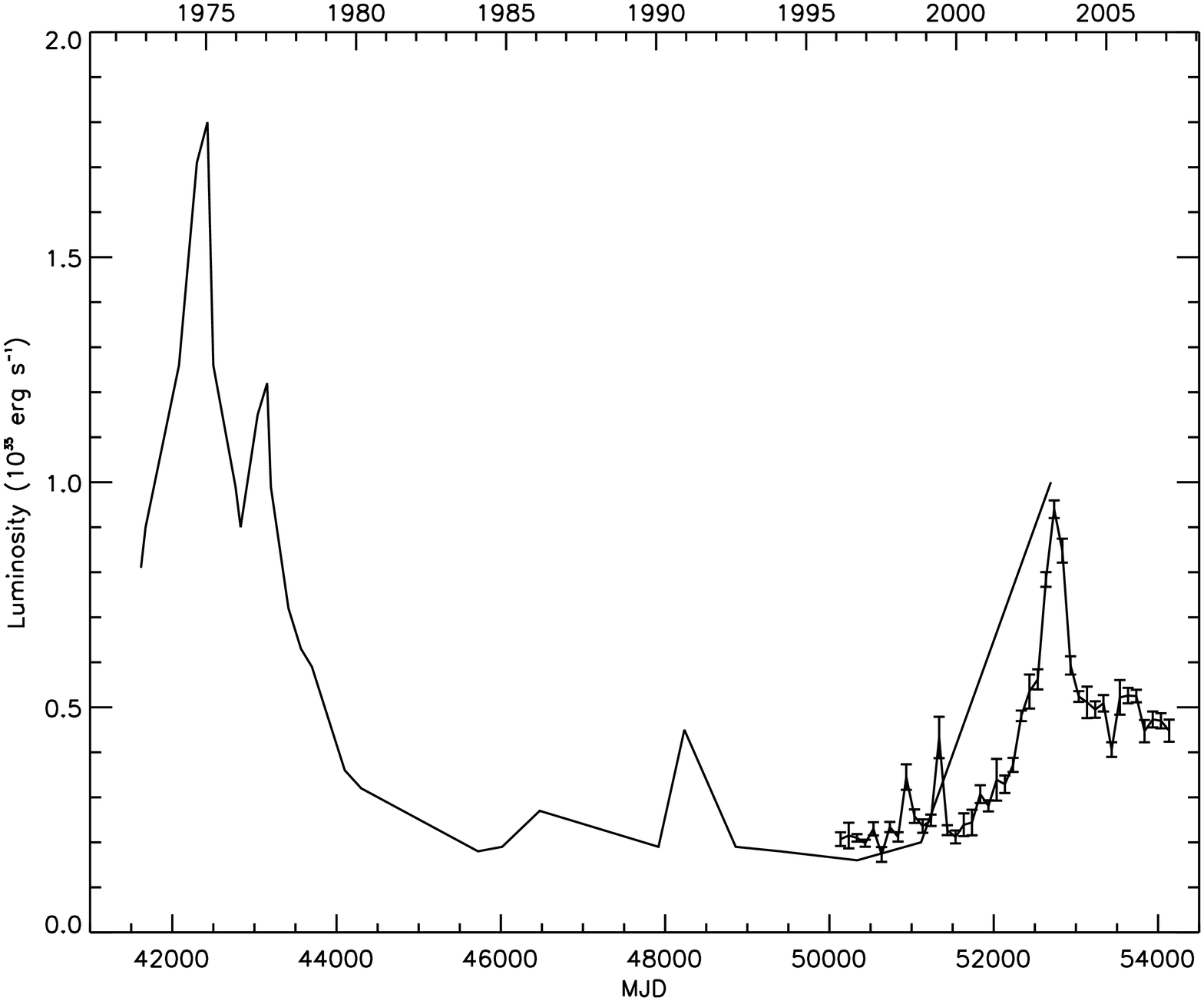}}
\resizebox{\hsize}{!}{\includegraphics[angle=0]{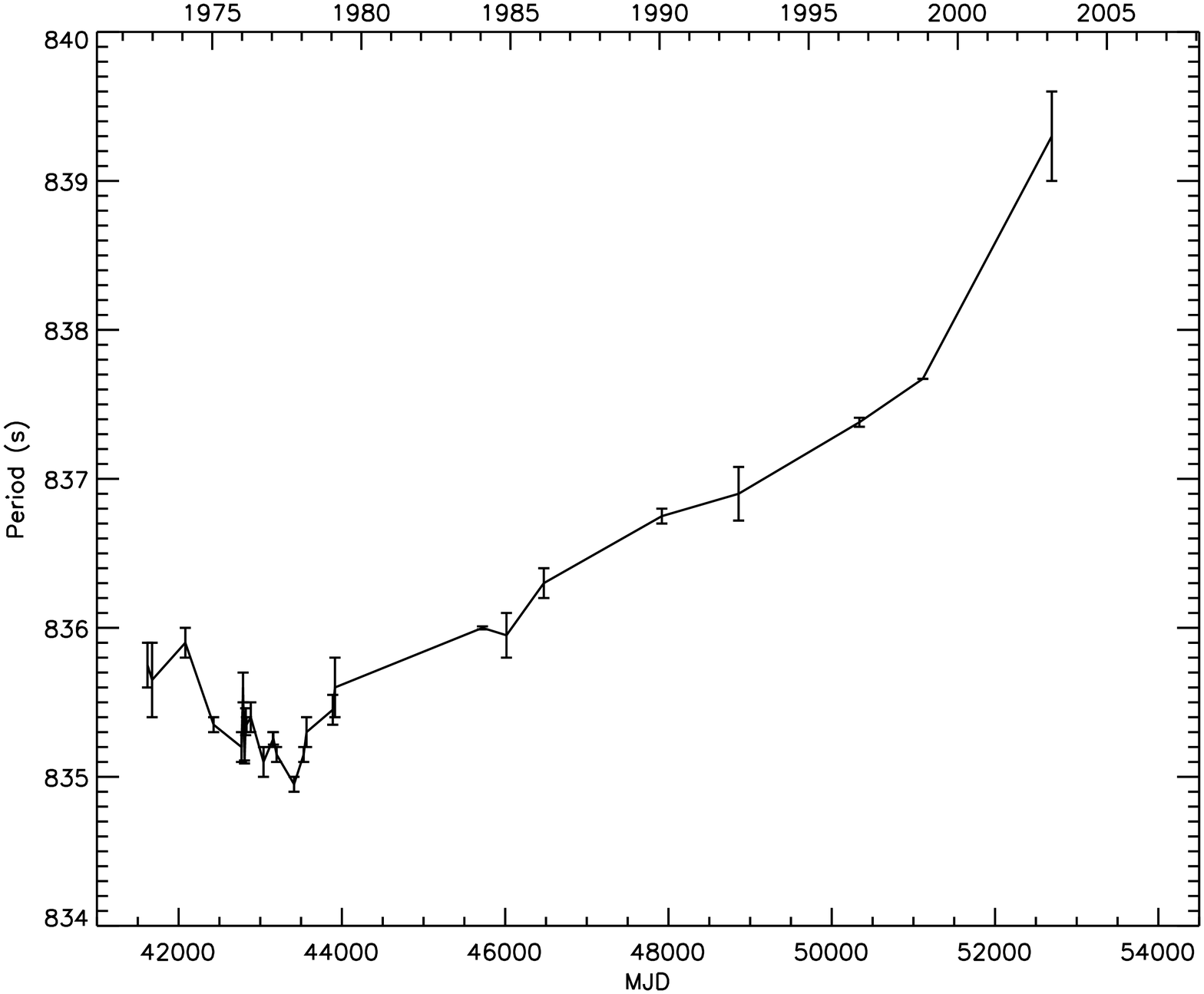}}
\caption{Luminosity (\textit{top}) and pulse--period (\textit{bottom}) history of \xper\ from 1972 October 29 to 2007 February 4. The luminosities of the top panel are based on the unabsorbed flux in the 2--10 keV and on a source distance of 1 kpc \citep{Telting+1998}. The data with error--bars are based on \XTE/\ASM\ measurements, which are reported for comparison.}\label{luminosity_period}
\vspace{-0.5 cm}
\end{figure}

In Fig.~\ref{luminosity_period} we report the long term evolution of \xper\ luminosity and spin period since the time of the first observations. The flux measured by \XMM~corresponds to $\sim 10^{35}$ erg s$^{-1}$, which is the highest level of luminosity after the end of the outburst in 1975. This result is confirmed also by the \XTE/\ASM\ data, which are reported for comparison: they clearly indicates that the \XMM~observation was performed just at a luminosity peak of the source, which was preceded by a steep increase and followed by a sharp decrease. This luminosity increase should imply a corresponding increase in the accretion rate and, in the case of matter transfer to the NS via an accretion disk, a larger momentum transfer to the neutron star, therefore we would expect to observe a pulsar spin--up. This is not confirmed by the measured pulse period, which is greatest than in all the previous observations. The pulse period evolution shows that the pulsar spin--down has proceeded also during the latest years, in spite of the luminosity increase. This is an independent evidence that the matter is transfered from the Be companion to the NS without forming an accretion disk.

The analysis of the pulse profile between 0.15 and 10 keV shows that, contrary to previous findings \citep[e.g.:][]{Robba+1996,DiSalvo+1998,Coburn+2001}, at all energies the pulse shape is far from being sinusoidal. It has a rich structure, with a main and a secondary peak, and varies with the energy. As a consequence, the peak of the hardness ratio between the count rates above and below 2 keV is not aligned with the maximum of the total count rate. It is possible that this change in the pulse profile is due to the source luminosity level, which was higher than in all the previous detections. The analysis of the source status around the pulse minimum provided another interesting result: in this phase interval also the hardness--ratio (2-4)/(0.15-2) reaches its minimum value, while the hardness--ratio (4-10)/(2-4) has a narrow peak. This behaviour was already observed by \citet{Robba+1996} and \citet{Coburn+2001} on \Ginga\ and \XTE\ data, respectively, at lower luminosity levels (i.e. 2--5$\times 10^{34}$ erg s$^{-1}$) and indicates a complex spectral evolution.

From the spectral point--of--view, we estimated a hydrogen column density N$_{\rm H}$ = (3.42$_{-0.08}^{+0.30}$)$\times10^{21}$ cm$^{-2}$, which is more than twice the value of $(1.51\pm0.22)\times10^{21}$ cm$^{-2}$ measured by Di Salvo et al. (1998) on \SAX\ data. However, we checked that this difference is due to the different fit model they used (i.e. a high--energy cut--off power--law). For reference, we note that optical observations give \textit{E(B-V)} = 0.395 \citep{Telting+1998}; assuming A$_{\rm V}$ = 3.1 \textit{E(B-V)} and the average relation A$_{\rm V}$ = N$_{\rm H} \times 5.59 \times 10^{-22}$ cm$^{-2}$ between optical extinction and X--ray absorption \citep{PredehlSchmitt1995}, this would predict N$_{\rm H}=2.19\times10^{21}$ cm$^{-2}$.

It is interesting to compare our spectral results with those obtained by \citet{Coburn+2001} on \XTE\ data, since they fit the source spectrum with a two component model equal to our model. They estimated a steeper power--law component ($\Gamma$ = 1.83 $\pm$ 0.03 instead of 1.48 $\pm$ 0.02), while the black--body component had a comparable temperature (kT = 1.45 $\pm$ 0.02 keV instead of 1.42$_{-0.02}^{+0.04}$ keV). However, in our case the black--body radius is much larger (361 m instead of 130 m): this is not surprising, since during the \XMM\ observation \xper\ was about 5 times more luminous than during the \XTE\ observation. In fact, in spite of this luminosity difference, also in the case of the \XTE\ observation $\sim$ 60 \% of the total flux was due to the power--law component.

As in previous observations of \xper, we found no evidence of an Fe--K line in the source spectrum. We estimated a 90 \% confidence upper limit on the equivalent width of almost 0.1 keV, i.e. an order of magnitude larger than in other works. On the other hand, we found a clue of an emission line at 0.58 keV, which was never found before. It is rather weak (EQW = 97 eV) and was not detected in all the pulse--phases, therefore its analysis requires further investigations.

Also \citet{Coburn+2001} performed a phase--resolved spectroscopy, even if with a different selection of the phase--intervals. Contrary to our findings, in their case both the power--law photon--index and the black--body temperature are almost constant along the pulse profile, and only the component normalizations change significantly. However, these results were obtained with a much lower count statistics and at a luminosity level about 5 times lower than in our case; furthermore, a fixed, common value of the hydrogen column density was forced to the different spectral phases and only data above 3 keV were considered.

The results we have obtained for \xper\ are in agreement with what has been observed in other two Be/NS binaries with long pulse period, i.e. in the low luminosity system \RX\ \citep{LaPalombaraMereghetti2006} and during the low--luminous quiescent state of the transient source \TreA\ \citep{Orlandini+2004,MukherjeePaul2005}. Also in these cases a thermal excess was detected above the main power--law: it has been modeled as a black--body with kT$_{\rm BB} >$ 1 keV and R$_{\rm BB} <$ 1 km, which contributes $\sim$ 30 \% of the total source luminosity. From this point of view, this class of low luminosity, long period Be/NS binaries is significantly different from the standard high--luminosity X--ray pulsars, whose thermal component is usually characterized by temperatures of $\sim$ 0.1 keV and radius of a few hundred km \citep[see][ for a review]{LaPalombaraMereghetti2006}.

\citet{Hickox+2004} showed that, in the low luminousity binary pulsars ($L_{\rm X}\le10^{36}$ erg s$^{-1}$), the thermal component can be due to either emission by photo--ionized or collisionally heated diffuse gas or thermal emission from the surface of the neutron star. Based on our results, the first option can be rejected, since we cannot reconcile it with the observed black--body spectrum. We therefore favor the interpretation of the soft excess in \xper\ as thermal emission from the neutron star polar cap. If we assume that the source is in the `accretor' status, with matter accretion on the NS surface, the black--body emitting radius of $\sim$ 360 m is consistent with the expected size of the polar cap. In fact, if M$_{\rm NS}$ = 1.4 M$_{\odot}$ and R$_{\rm NS}$ = $10^6$ cm, the source luminosity of $1.4\times10^{35}$ erg s$^{-1}$ implies an accretion rate $\dot M = 7.5 \times 10^{14}$ g s$^{-1}$ and, adopting B$_{\rm NS} = 10^{12}$ G, a magnetospheric radius R$_m = 8.4 \times 10^8$ cm \citep{Campana+1998}. In this case, based on the relation R$_{col}\sim$ R$_{\rm NS}$ (R$_{\rm NS}$/R$_m$)$^{0.5}$ \citep{Hickox+2004}, we would obtain R$_{col} \sim$ 345 m, i.e. a value in good agreement with the estimated black--body emitting radius.

Based on the above result, it makes sense to attribute the observed thermal component to the emission from the NS polar caps, as in the case of \RX\ and \TreA. If this description is correct, we would expect to observe some variability of the soft component along the pulse phase. From this point of view the phase--resolved spectroscopy provides no conclusive results, since it proves that both a variable and a constant thermal component can account for the observed spectral variability.

\vspace{-0.3 cm}
\section{Conclusions}

Analyzing a $\sim$ 31 ks long \XMM\ archival observation of the Be/NS binary system \xper, we found that the source was detected at a luminosity level of $\sim1.4\times10^{35}$ erg s$^{-1}$, which is the highest value ever reported since the large outburst that occured on 1975. This result was independently confirmed by the \XTE/\ASM\ data, which has been monitoring the source during the latest decade.

The timing analysis confirmed the pulsar spin--down already detected in the previous observations. Moreover, it shows that the pulse profile has a complex structure and is energy dependent. 

From the spectral point--of--view, thanks to the high effective area of \XMM\ over a wide energy band, we could perform an accurate analysis also at low energies. In this way, we detected a data excess above the main power--law component, which contributes for $\sim$ 40 \% to the total source luminosity. Based on its black--body spectrum, with an emission radius of $\sim$ 360 m, this component could be attributed to a polar--cap emission. It is interesting to note that the same hypothesis has been advanced also to describe the spectra of other two low--luminosity and long--period Be/NS pulsars, i.e. \RX\ and \TreA\ (during its quiescent state).

We found no evidence of a narrow Iron K$_{\alpha}$ line between 6 and 7 keV; on the other hand, our data suggest the presence of an emission feature at $\sim$ 0.58 keV, whose description and nature require further investigations.

The phase--resolved spectroscopy shows a large spectral variability along the pulse period, which can be modeled with both a variable and a constant thermal component. This result supports but does not confirm the polar--cap origin of this emission, therefore further studies will be necessary in the future in order to solve this issue.

\begin{acknowledgements}
This work is based on observations obtained with {\em XMM--Newton}, an ESA science mission with instruments and contributions directly funded by ESA Member States and NASA. The {\em XMM--Newton} data analysis is supported by the Italian Space Agency (ASI), through contract ASI/INAF I/023/05/0.
\end{acknowledgements}

\vspace{-0.5 cm}
\bibliographystyle{aa}
\bibliography{biblio}

\end{document}